# Speckle Imaging with VLT/NACO No-AO Mode

Sridharan Rengaswamy[1]
Julien Girard[1]
Willem-Jan de Wit[1]
Henri Boffin[1]

[1] ESO

Long-exposure stellar images recorded with large ground-based telescopes are blurred due to the turbulent nature of the atmosphere. The VLT employs active and adaptive optics (AO) systems to compensate for the deleterious effects of the atmosphere in real time. The speckle imaging technique provides an alternative way to achieve diffraction-limited imaging by post-processing a series of short-exposure images. The use of speckle imaging with the no-AO mode of NACO at the VLT is demonstrated. Application of this technique is particularly suited to the *J*-band and it provides versatile high angular resolution imaging under mediocre conditions and/or in imaging extended objects. The implementation of this mode underlines the continuing attractiveness of NACO at the VLT.

## Astronomical speckle imaging

The invention of speckle interferometry (Labeyrie, 1970) has revolutionised the field of high-resolution imaging with ground-based telescopes. In speckle interferometry, a series of short-exposure images of the object of interest (for example, a spectroscopic binary star) is recorded with a spectral filter whose bandwidth is smaller than the mean wavelength (quasi-monochromatic conditions). The modulus-squared Fourier transform of the images averaged over the ensemble (a quantity generally referred to as the energy spectrum) preserves the information up to the diffraction limit of the telescope. Deconvolving this average with a similarly obtained average of an isolated single star, close in time and angular distance to the target, removes the effects of the atmosphere and the telescope on the energy spectrum, leaving behind the energy spectrum of the object of interest. A subsequent Fourier transform of the energy spectrum provides the diffraction-limited autocorrelation of the object of interest. In the case of binary stars, the binary nature can be visually identified from the peaks of the autocorrelation.

Speckle interferometry has made significant contributions to the field of binary star research by creating several binary star catalogues (e.g., Mason et al., 2013). Since the energy spectrum does not preserve the Fourier phase information, it is not possible to obtain an image of the object apart from the autocorrelation. Hence there is a 180-degree ambiguity in the position angle of the binary star. In 1977, Gerd Weigelt showed, experimentally, that the phase of the complex triple product of the Fourier transform of the images at three spatial frequencies (the third frequency being the sum of the first two) averaged over the ensemble, provides the sum of the Fourier phases of the object at those three frequencies. This complex product, called the bi-spectrum, is immune to atmospheric turbulence (as the random phase errors introduced by the atmosphere are completely cancelled in the bi-spectrum) and provides a way to retrieve the Fourier phases of the object.

It should be noted that the phase of the bi-spectrum is the same as the "closure phase", a term coined by Jennison (1958) in radio astronomy and used first by Rogstad (1968) for optical imaging through turbulence. Combining the phase information with the energy spectrum obtained from the speckle interferometry, an image of the object of interest can be reconstructed. This technique is known as speckle imaging.

## The no-AO mode of NACO

The Very Large Telescope (VLT) instrument NAOS–CONICA (NACO; Lenzen et al., 2003) was built at the end of the 1990s and its deformable mirror has a total of 185 actuators (Rousset et al., 2003). The no-AO mode of NACO (Girard et al., 2010), essentially bypasses the adaptive optics module NAOS and uses CONICA's "burst mode" to record a series of short-exposure images, exploiting the detector's windowing capability. There are different ways in which these short-exposure images can be processed. Holographic imaging of crowded fields was pioneered and presented by Rainer Schödel (Schödel & Girard, 2012; Schödel et al., 2013). Here we demonstrate how the NACO non-AO mode, coupled with the aforementioned speckle image processing technique, can be exploited to obtain near diffraction-limited images of objects of interest. We have developed the required pipeline data reduction tool (Rengaswamy, Girard & Montagnier, 2010) for reconstructing an image from the speckle data[1].

## Imaging binary stars

The study of binary stars is an important branch of astrophysics. It is the most common way by which stellar masses are determined. Multi-epoch images of a binary system enable the calculation of the orbital elements of the binary. A typical long-exposure image does not resolve very close binary stars, i.e., those at sub-arcsecond separation, as the resolving power is impaired by the atmospheric turbulence. Short-exposure images (freezing the atmosphere) can be post-processed to resolve the binary. The bottom left panels of Figure 1 show images of the binary star HIP24800, reconstructed from a series of 900 speckle images recorded in NACO no-AO mode in January 2010. The top panels show the best and the worst images of the series. The best image is identified as the one for which the sum of the fourth power of the mean subtracted pixel intensities is the highest. The bottom right panel shows the long-exposure image, a simple mean of all the short exposures. The separation between the components of the binary is 161 milliarcseconds (mas) and their brightness ratio is about 2.3.

## Imaging near-Earth objects

The no-AO mode of NACO also makes the imaging of fast-moving near-Earth objects at high angular resolution feasible. The near-Earth passage of the asteroid 2005 YU$_{55}$ in November 2011 provided an excellent opportunity to test this. As the adaptive optics could not lock onto this fast-moving object, data were recorded in the no-AO mode in the *Ks*-band. Figure 2 shows the resulting average reconstructed image of the asteroid. The nominal angular resolution of 60 mas at the Ks-band corresponds to a linear resolution of 95 metres at the distance (329 900 km) of the asteroid. The dimensions of the asteroid were much larger than this resolution and thus the technique allowed the asteroid to be spatially resolved. The cross-sections of the image, subjected to an edge enhancement operator (Prewitt edge enhancement operator of the IDL software) to retrieve the asteroid dimensions, indicated a size of (261 ± 20) × (310 ± 30) metres. For comparison, tri-axial diameters of 337 × 324 × 267 metres, with uncertainties of 15 metres in each dimension, were estimated for the size of the asteroid from the Keck adaptive optics system, six hours before our observations. Accounting for the time difference, and the 18-hour rotation period of the asteroid, our results are in good agreement with those of the Keck AO system, despite the mediocre observing conditions (seeing of 1.2 to 1.5 arcseconds at 550 nm and atmospheric coherence time of two milliseconds). These observations demonstrate the robustness of the no-AO mode and the speckle imaging technique.

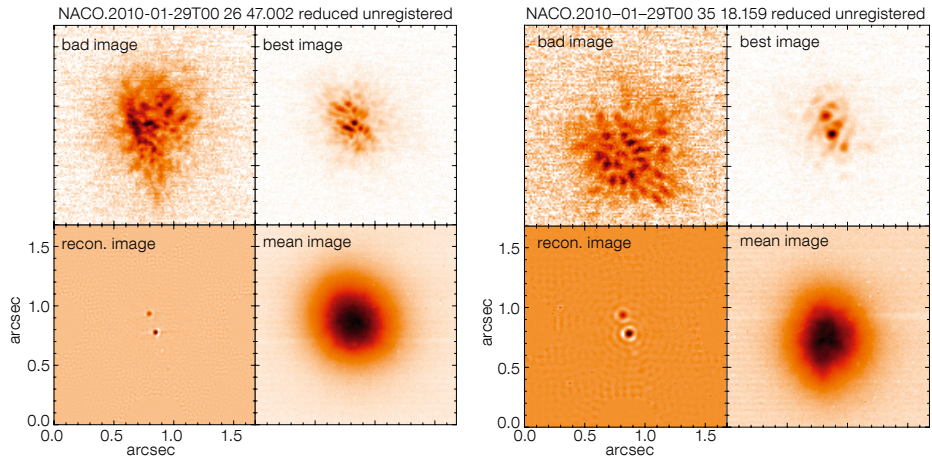

Figure 1. Speckle reconstructed images based on NACO no-AO observations of the binary HIP 24800 in J- (left) and Ks- (right) bands (low left panels). The top panels show the worst and best images of the recorded series. The lower right panels show the mean image.

### Imaging circumstellar envelopes around evolved stars

L2 Pup (HD 56096, HR 2748) is a semi-regular pulsating red giant star in the constellation of Puppis with an angular diameter of 17 milliarcseconds. Figure 3 shows speckle-reconstructed images of L2 Pup at 2.27 μm observed with the no-AO mode. The nominal resolution at this intermediate band is about 60 mas. While the stellar continuum is unresolved, the extended circumstellar emission from this star is visible in the reconstructed images. We have suppressed the features with intensity values less than 5% of the maximum intensity in generating this figure. The circumstellar envelope appears enlarged in the second reconstruction, obtained six months after the first one. This is perhaps due to the 140-day pulsation period of the star or due to its high mass-loss rate. The panel on the left is the ideal point spread function at 2.27 μm. The middle and right panels indicate speckle reconstructions obtained from the March and September 2012 data, respectively. These images serve first and foremost as extremely valuable information on the dynamical evolution of the circumstellar envelope at high angular resolution. Secondly, they can serve as a good starting point for even higher angular resolution investigations with, for example, the Very Large Telescope Interferometer.

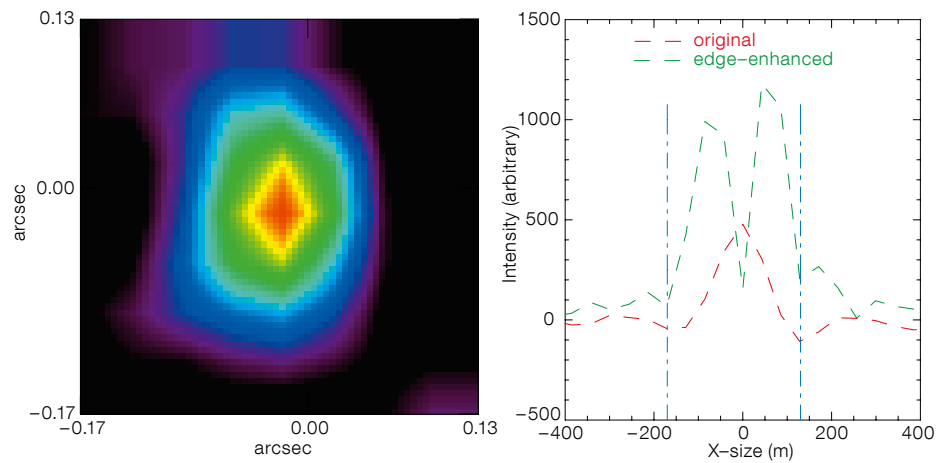

Figure 2. Reconstructed image of asteroid 2005 YU$_{55}$ (left) and its horizontal cross-section (right). The red line indicates the cross-section before applying an edge detection operator to the reconstructed image. The green line indicates the cross-section after applying the edge detection operator. The blue lines indicate the size.

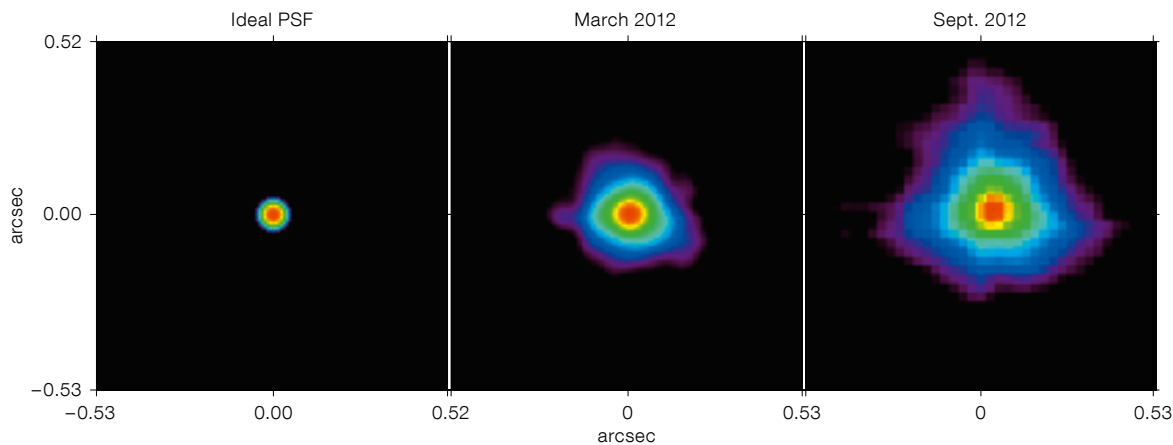

Figure 3. Speckle image reconstruction of the circumstellar envelope of L2 Pup, observed in March and September 2012. The theoretical (ideal) PSF is shown for comparison.

Allen et al. (1972) alluded to the possible presence of dust shells around carbon-rich Wolf–Rayet stars (WC stars) based on the near-infrared excess inferred from photometric observations. This was initially a surprising result, as the dust particles would be expected to be destroyed by the strong stellar winds from these stars. Therefore, these observations indicated the continuous formation of dust somewhere within the environment of these stars. With the advent of aperture masking interferometry, direct imaging of nearby Wolf–Rayet stars became feasible. A dusty pinwheel nebula structure was discovered around two persistently dust-forming WC stars, *viz.* WR 104 (WC9d + B0.5V) and WR 98a (WC8-9vd) (Monnier et al., 1999; Tuthill et al., 1999). Although these systems could be explained if the dust can be formed at the downstream regions of the shockfront formed by the colliding winds of the WR star and an associated secondary star of type OB, this has never been conclusively proven.

In an effort to clarify whether the dust formation in WC stars occurs at the colliding wind region or in the self-shadowing regions of the clumpy winds of an isolated star, we executed a pilot survey of six WC stars, consisting of three putative binaries and three isolated single stars. Figure 4 shows the speckle-reconstructed images of WR 113 and WR 69 (supposedly binaries) and WR 15 (a single star) in the *JHKs*-bands. The circumstellar envelope shows extended emission confirming the presence of the dust. By resolving the dust emission in several WC stars, at multiple epochs in *J*-, *H*-, *K*- and *L′*-bands, one can trace the temperature of the dust distribution and thus distinguish between the two scenarios of the dust formation.

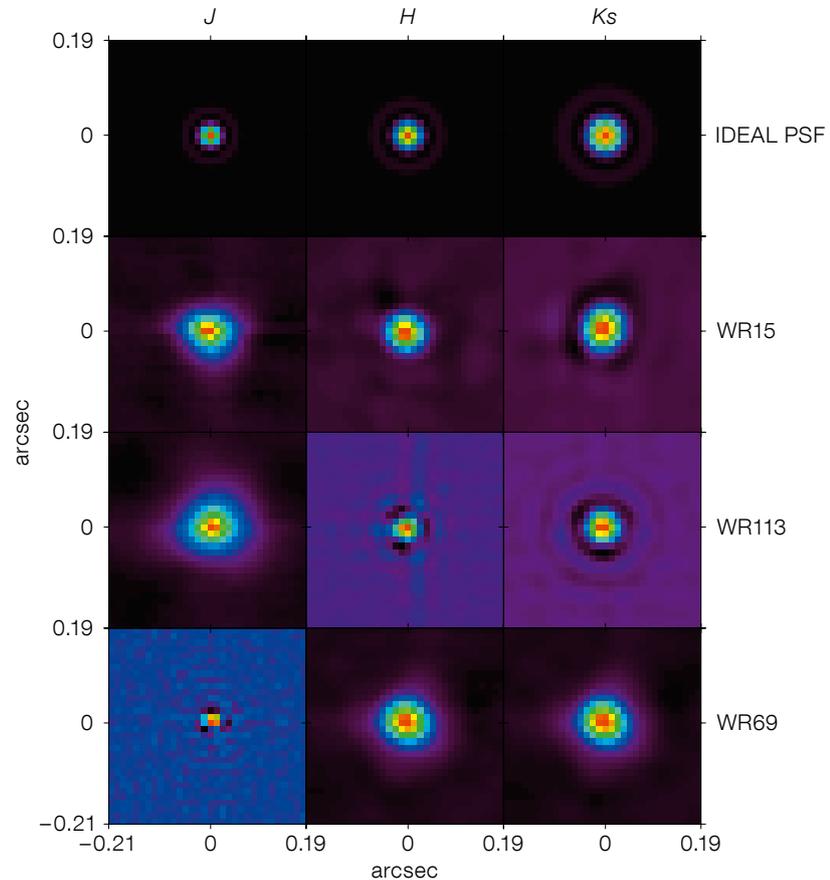

Figure 4. Reconstructed images of WR 15, WR 113 and WR 69 in *J*-, *H*- and *Ks*-bands. The top row shows theoretical (ideal) point spread functions in *J*-, *H*- and *Ks*-bands.

## Comparison between speckle-reconstructed and AO-corrected images

In an effort to assess the performance of speckle imaging, AO-corrected images were recorded nearly simultaneously along with the speckle datacubes in the case of the binary star shown in Figure 1. Figure 5 shows good comparison between the speckle-reconstructed images obtained with the NACO no-AO mode and the corresponding AO-corrected images in *J*- (left panel) and *Ks*- (right panel) bands. The speckle-reconstructed images have been rescaled (between 0 and 255) and the contrast has been reversed for the sake of display. Qualitatively, they match very well.

In order to have a quantitative comparison, we estimated the Strehl ratios of the binary components from the reconstructed images. It should be noted that speckle reconstruction yields the "object" intensity distribution on the sky (of course sampled at the spatial sampling of the detector). This is quantitatively different from the "image" intensity distribution (i.e., the object convolved with the response of the telescope) or the point spread function (PSF) obtained in normal or adaptive optics imaging. Thus, prior to estimating the Strehl ratios, we convolved the speckle reconstructions with an ideal synthesised PSF of the telescope. We estimated the Strehl ratio as the ratio of volume under the transfer function (Fourier transform of the PSF) of the reconstructed image (binary components in this case) to the volume under the transfer function of an ideal (synthesised) PSF. Strehl ratios were obtained with a window size of 16 × 16 pixels.

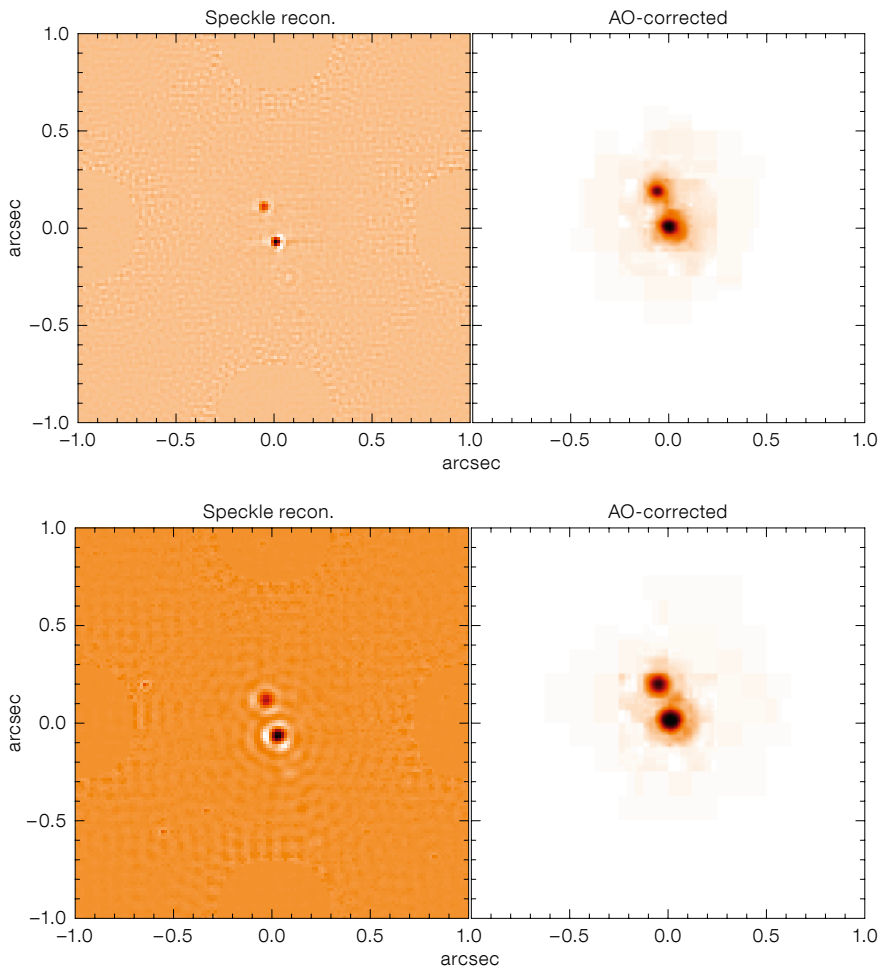

Figure 5. Comparison between speckle-reconstructed images and the images obtained with AO corrections at the telescope in *J*- (upper) and *Ks*- (lower) bands.

The average Strehl ratios of ten consecutive measurements in *J*- and *Ks*-bands were about 70 % and 90 % respectively under very good conditions (0.5–0.6 arcseconds seeing) observing close to the zenith. We clearly see that speckle imaging tends to provide a higher Strehl ratio than AO (typically 60 % Strehl in *Ks*-band and less at shorter wavelengths), under good conditions. This could be interpreted as follows: as speckle image reconstruction "corrects" for the phase aberrations up to infinite order (unlike adaptive optics imaging, which corrects the phase aberrations up to a finite order of equivalent Zernike modes), we obtain higher Strehl ratios than adaptive optics imaging or normal imaging.

### When to use speckle imaging with the no-AO mode

It is important to consider when to use NACO no-AO instead of AO. In general, speckle imaging with the no-AO mode is more beneficial in the *J*-band where the AO fails to perform well due to its limited number of wavefront sensing/correcting elements. Further, it may not be always possible to lock the AO loop on fast-moving, extended, near-Earth objects even under favourable conditions (since AO performs best on point sources) and speckle imaging is extremely valuable for imaging such objects (e.g., asteroid 2005 YU$_{55}$).

AO stands out gloriously when it is required to image faint targets with long exposures (deep imaging). It minimises confusion in crowded fields (e.g., the Galactic Centre) by separating close sources. Its use is inevitable for high-resolution spectroscopy and should be preferred when the observing conditions are excellent. However, there could be situations where AO cannot be used efficiently. For example, when an object is very bright (e.g., L2 Pup with *Ks* = –2.3) long exposures will saturate the detector and thus short exposures with the no-AO mode should be preferred.

Another advantage of speckle imaging is the possibility of excluding bad images that correspond to highly turbulent states of the atmosphere. In other words, one can select the best images from the recorded series of images and then obtain a speckle-reconstruction from them. This is in line with the concept of "lucky imaging", where typically 10 % of the best images are used with a simple shift-and-add method to obtain a high-resolution image.

In our speckle processing, we use 90 % of the frames when all images in the recorded sequence have similar contrast. However we discard more than 50 % of the frames when the contrast of the frames shows significant variations, which occurs particularly under mediocre conditions. This flexibility facilitates speckle imaging under a wide variety of conditions. For example, the reconstructions of L2 Pup were obtained from the data recorded under 1.2–1.6 arcsecond seeing and one millisecond coherence time, after discarding nearly 80 % of the frames, thanks to the ability to record a few thousand frames. But the AO loop will not be stable under highly varying conditions. Further, the quality of the AO correction may not be good under unstable atmospheric conditions, while it would still be possible to obtain a speckle reconstruction after excluding frames with degraded image quality.

### Additional benefit of the No-AO mode

An additional use of this mode is the ability to ascertain the proper tracking of the telescope (under wind shake, vibrations and occurrences of the primary mirror [M1] passive support motions). A burst of 8000 images of a bright target, each of 20 mas exposure spans 160 seconds. A software-based image registration process (cross correlation using Fourier transform [FCC]) can generate a plot of image motion as a function of time (as shown in Figure 6). Any slowly varying jitter with large amplitude (a few arcseconds) could be attributed to the residual tracking errors of the telescope. The estimated root mean square (rms) jitter for the case shown in Figure 6 is about 113 mas over 2.5 minutes with peak-to-peak jitter of about 0.6 arcseconds. There were no passive support motions in M1 which could momentarily increase the amplitude of the jitter. The estimated jitter is compatible with the expected (as per design specifications) rms tracking error for the VLT Unit Telescopes of 100 mas over 30 minutes.

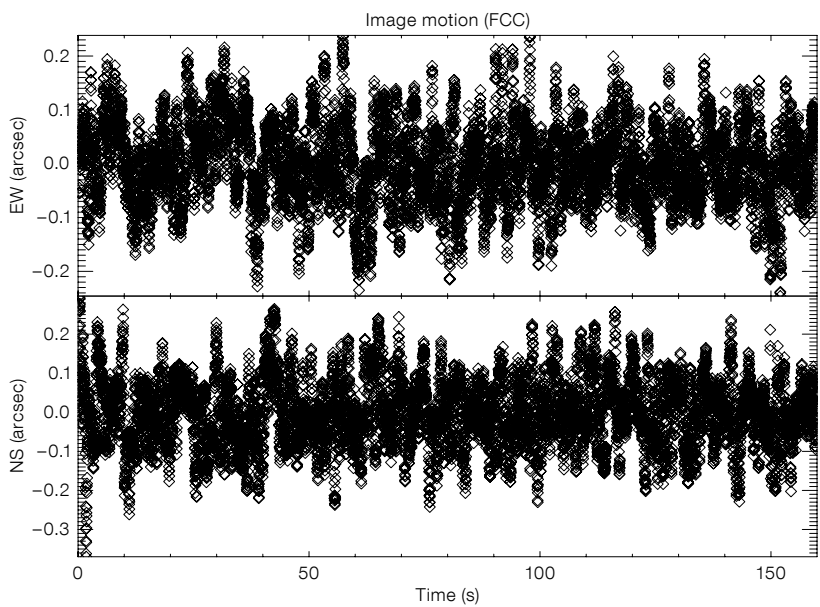

Figure 6. Image jitter estimated by cross-correlating the speckle frames with a reference frame. The rms jitter can be used to assess the quality of the tracking. Wind shake and the M1 passive support motions can cause a momentary increase in the magnitude of the jitter.

### Links

[1] The speckle imaging data reduction tool can be obtained on request to: srengasw@eso.org

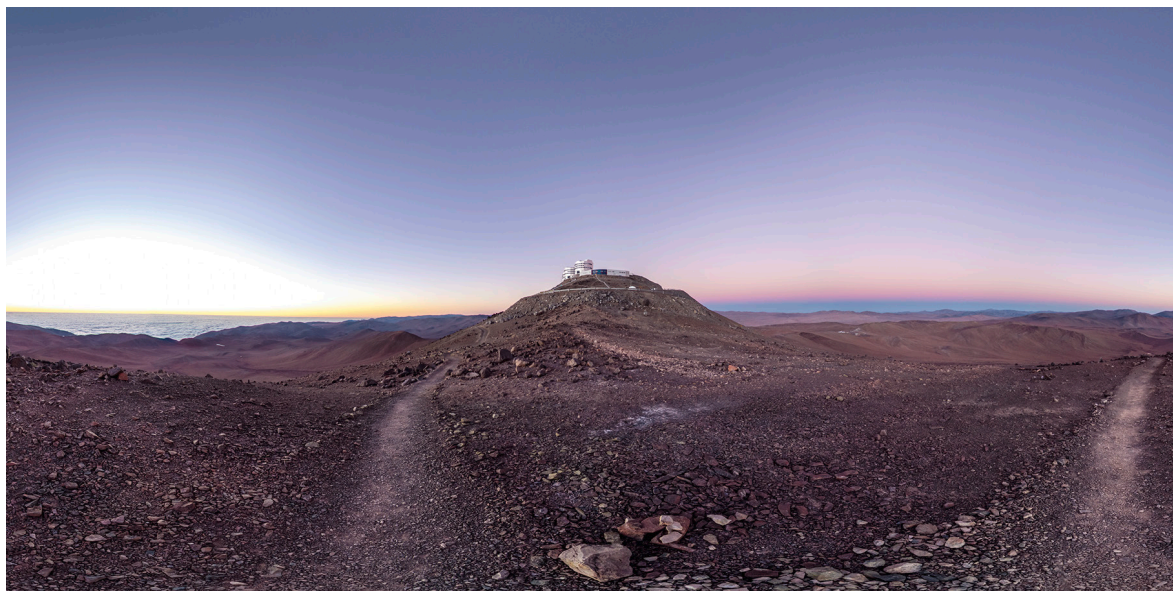

Sunset over Paranal on 5 July 2012, the date of the very low precipitable water vapor event as described in the following article. The photo was taken by ESO Photo Ambassador Gabriel Brammer, who found the scene to be extraordinarily clear and beautiful. Little did he know that the following night would be one of the driest on record. See Picture of the Week for 3 February 2014 for details.